\documentclass[prd,showpacs,showkeys,nofootinbib,superscriptaddress]{revtex4-1}
\usepackage{graphicx}
\usepackage{dcolumn}
\usepackage{epsfig}

\usepackage{bm}

\usepackage{latexsym,epsfig,amsmath,amssymb}
\usepackage{amssymb}
\usepackage{amsmath}
\usepackage{amsfonts}
\usepackage{epsfig} 
\usepackage{verbatim} 

\newcommand{\eq}{\begin{eqnarray}}

\newcommand{\en}{\end{eqnarray}}

\newcommand{\ba}[1]{\begin{eqnarray} \label{(#1)}}

\newcommand{\ea}{\end{eqnarray}}

\begin{document}

\title{Beta Decaying Nuclei as a Probe of Cosmic Neutrino Background}

\author{Amand Faessler}


\affiliation{Institute f\"{u}r Theoretische Physik der Universit\"{a}t 
T\"{u}bingen, D-72076 T\"{u}bingen, Germany}
\author{Rastislav  Hod\'ak}
\affiliation{Department of Nuclear Physics and Biophysics, 
Comenius University, Mlynsk\'a dolina F1, SK--842 15
Bratislava, Slovakia}
\author{Sergey Kovalenko}
\affiliation{Departamento de F\'{i}sica, 
Universidad T\'{e}cnica Federico Santa Mar\'{i}a\\
and Centro Cient\'{i}fico-T\'ecnologico de  Valpara\'{i}so\\
Casilla 110-V, Valpara\'{i}so,
Valpara\'{i}so, Chile}
\author{Fedor \v Simkovic}
\affiliation{Department of Nuclear Physics and Biophysics, 
Comenius University, Mlynsk\'a dolina F1, SK--842 15
Bratislava, Slovakia}
\affiliation{Laboratory of Theoretical Physics, JINR,
141980 Dubna, Moscow region, Russia}

\date{\today}

\begin{abstract}
We analyze the possibility of detection of cosmological relic neutrinos via neutrino 
capture on  beta decaying nuclei. 
This reaction has no  threshold in neutrino energy, which is crucial for searching for relic 
neutrinos possessing very low energies.
We focus on tritium ($^{3}$H) and rhenium ($^{187}$Re) beta radioactive isotopes to be  used 
in KATRIN and MARE experiments dedicated to measurement of the electron neutrino mass at sub-eV scale.
We examine these experiments  from the viewpoint of searching for the cosmological 
neutrinos via neutrino capture. We conclude that even with possible gravitational clustering of 
relic neutrinos the prospects for their detection in these and other similar experiments 
are not optimistic. Nevertheless KATRIN and MARE experiments could establish some usefull constraints
on density of relic neutrinos.
\end{abstract}



\maketitle

\section{Introduction}
\label{Introduction}
The cosmological relic neutrinos are one of the most important and abundant
constituents of the Universe predicted by  Big Bang 
cosmology.  
There are about $10^{87}$ neutrinos per flavor in the visible  Universe \cite{eidel}, which corresponds to
the number density per  flavor 
of relic \mbox{(anti-)neutrinos} in average over the Universe about 
$\langle\eta\rangle \sim 56~cm^{-3}$\cite{giunti}.
In number, neutrinos exceed the 
constituents of ordinary matter (electrons, protons, 
neutrons) by a factor of ten billion. 
On the other hand their existence has not yet been confirmed by direct observations. 
This represents a challenging  problem of modern cosmology
and experimental astroparticle physics. 
Many proposals in the literature aim to observation of  indirect astrophysical manifestations 
of relic neutrino sea \cite{astro1,astro2,astro3,astro4,astro5,astro6}. 
However laboratory experiments would be the most robust probe of  this component of the Universe.  
There are various strategies discussed for laboratory searches of relic neutrinos \cite{Weinberg}, 
\cite{lab1,lab2,lab3,lab4,lab5,lab6,lab7,cocco,vogel,Hodak:2009zz,Hodak:2011zz}.  The neutrino capture reaction, 
\cite{Hodak:2009zz}
\begin{equation}\label{nucapt10}
\nu_{e} + (A,Z) \rightarrow (A,Z+1)+ e^-,  
\end{equation}
is one of the most discussed possibilities in the literature 
\cite{Weinberg,cocco,vogel,Hodak:2009zz,Hodak:2011zz}.
Since relic neutrinos have very low energy this reaction should have no 
threshold implying that the initial nucleus is $\beta$-radioactive. This idea was 
first proposed many years ago by Weinberg in Ref. \cite{Weinberg}, but in the context 
of massless neutrinos  with large chemical potential.  The latter proved to be 
inconsistent with the Big Bang cosmology.  The discovery of neutrino mass $m_{\nu}$ 
revived the interest to the above reaction, since at the endpoint of the electron 
spectrum the energy difference between $\beta$-decay and neutrino capture electrons 
should be around  $\sim 2 m_{\nu}$. In principle, this effect can be observed with 
a detector having an energy resolution less than this difference.
However, the existing experimental techniques are unable to detect relic 
neutrinos via neutrino capture if their number density in our vicinity does not significantly differ from the global average number density $\langle\eta\rangle\sim 56~cm^{-3}$ \cite{vogel,Hodak:2009zz,Hodak:2011zz}. 
As it was pointed out recently \cite{clust,clust-det,vogel},  the gravitational 
clustering of neutrinos in our galaxy or galaxy cluster may drastically enhance 
the relic neutrino density in comparison to the overall average $\langle\eta\rangle$ making its detection more realistic. 

In our previous papers \cite{Hodak:2009zz,Hodak:2011zz} we studied reactions of neutrino capture on single (\ref{nucapt10}) and double beta decaying nuclei as a tool for probing cosmological relic neutrinos.
Here we focus on detailed calculation of the event rate of the reaction (\ref{nucapt10}) and estimate typical detection rates of the relic neutrino capture taking into account possible effect of gravitational clustering. 
In Sec. \ref{sec1}  we present our approach to nuclear $\beta$-decay with an emphasis on 
tritium $^{3}$H and rhenium $^{187}$Re radioactive isotopes. In Sec. \ref{sec2} we apply these approach to analysis of relic neutrino capture on these isotopes and discuss KATRIN $^{3}$H  \cite{katrin,otten} and  MARE $^{187}$Re experiments \cite{mare} dedicated to measurement of electron neutrino mass at sub-eV scale. We examine prospects of these experiments for searching for the relic neutrinos via neutrino capture on the corresponding nuclei.

\section{Nuclear $\beta$-decay. Basic ingredients}
\label{sec1}

Here we discuss theoretical aspects of single $\beta$-decay of nuclei with special emphasis on 
tritium and rhenium isotopes  \cite{Rhen}. We introduce nuclear structure ingredients used in the subsequent sections for analysis of relic neutrino capture by these nuclei.

Let us consider general case of nuclear  $\beta$-decay
\begin{equation}\label{beta-dec1}
(A,Z) \rightarrow (A,Z+1)+ e^- + \overline\nu_{e}\, ,  
\end{equation}
driven
 by the standard weak $\beta$-decay Hamiltonian
\begin{equation}
{\cal H}^\beta(x)=\frac{G_{\beta}}{\sqrt{2}} 
\bar{e} (x)\gamma^{\mu} (1- \gamma_5) \nu_{e}(x)
j_\mu (x) + {h.c.}.
\label{eq.2} 
\end{equation}
Here, $G_\beta = G_F \cos{\theta_C}$, where $\cos{\theta_C}$ is the Cabbibo angle. 
$e(x)$ and $\nu_{{e }}(x)$ 
are the electron and neutrino fields, respectively. 
The strangeness conserving free nucleon charged current is
\begin{equation} 
j^{\mu} (x)=\bar{p}(x)\gamma^{\mu} (g_{{V}}-
g_{{A}} \gamma_{{5}}) n(x),
\end{equation}
where $p(x)$ and $n(x)$ are the proton and neutron fields, respectively.
$g_{{V}}=1.0$ and $g_{{A}}=1.25$.

Single $\beta$-decay occurs in the first order in the weak interaction.  
The corresponding S-matrix element is given by:
\begin{eqnarray}\nonumber
\langle{f}|S^{{(1)}}|{i}\rangle&=& 2\pi \delta(E_{f} + E_e + E_\nu - E_{i}) \langle {f}|T^{{(1)}}|{i}\rangle  =\\
\nonumber
&=& 2\pi\delta (E_f + E_e + E_\nu - E_i)
(-i) \frac{G_{\beta}}{~\sqrt{2}} \times \nonumber\\
\label{ME1}
&& \int   
{\bar{\psi}}_e (\bm{x},E_e ) \gamma^{\mu} (1-\gamma_{{5}}) \psi^c_\nu (\bm{x}, E_\nu ) 
\langle A,Z+1| J_{\mu} (0,\bm{x})|A,Z\rangle   d\bm{x}.
\end{eqnarray}
Here, $J_{\mu}$ is the nuclear weak charged current in the Heisenberg representation. In our notation
$E_{i}$, $E_{f}$, $E_{e}$ and $E_{\nu}$ are energies of initial and final 
nuclei, electron and antineutrino, respectively. 
The wave functions of outgoing electron and antineutrino are denoted as 
$\psi_{e}, \psi^c_{\nu}$. We use non-relativistic impulse approximation for the
nuclear current:
\begin{equation}\label{NRIA}
J^{\mu} (0,\bm{x}) = \sum_{n=1}^{A} \tau^+_n [g_V g^{\mu 0} + g_A  (\sigma_k)_n g^{\mu k} ] \delta(\bm{x}-{\bm{x}}_n).
\end{equation}
From now on it is convenient to consider $\beta$-decay of tritium and rhenium separately. 

\subsection{$\beta$-decay of tritium}
\label{sec11}

In the tritium $\beta$-decay the spin and parity of initial and final nuclei are equal:
\begin{equation}
\label{Tr-b}
{^{3}H}((1/2)^+) \rightarrow {^{3}He}((1/2)^+) + e^- + {\overline{\nu}}_e, 
\end{equation}
In this case the dominant contribution to the decay rate 
is determined by $s_{1/2}$  wave-states of outgoing electron and  antineutrino
\begin{eqnarray}\label{WFS1e}
\psi_e({\bm{x}}, E_e) &\approx& \sqrt{F_0(Z+1,E_e)}~ u(P_e) 
\\
\label{WFnu}
\psi^c_\nu({\bm{x}},E_\nu) &\approx&  u(-P_\nu) 
\end{eqnarray}
where $\bm{x}$ is the coordinate of the lepton. $F_0$ is the relativistic Fermi 
function (for more details see Appendix) \cite{doi}. 
Normalization of the spinor is $u^\dagger(P)u(P) = 1$.
$P_e\equiv (E_e, {\bm{p}}_e)$ and  $P_\nu\equiv (E_\nu, {\bm{p}}_\nu)$
are 4-momenta of electron and antineutrino, respectively.

Substituting Eqs. (\ref{WFS1e}), (\ref{WFnu}) and  (\ref{NRIA}) into Eq. (\ref{ME1})  we get 
for T-matrix element an expression
\begin{eqnarray}\label{T1}
\langle {f}|T^{(1)}|{i}\rangle  &=& (-i) 
\frac{G_{\beta}}{~\sqrt{2}} \sqrt{F_0(Z+1,E_e)} ~ \times \nonumber\\
&&\bar{u}(P_e) \gamma_{\mu} (1-\gamma_{{5}})u(-P_\nu ) 
\left[ g^{\mu 0} M_F + g_A g^{\mu k} (\bm{M}_{GT})^{k}
\right],
\end{eqnarray}
where Fermi and Gamow-Teller nuclear matrix elements are defined as
\begin{eqnarray}
M_F(m,m') &=& {_{{^3}He}\langle }(1/2)^+ m'| \sum_n \tau^+_n |(1/2)^+ m{\rangle _{{^3}H}}, \nonumber\\
\bm{M}_{GT}(m,m') &=& {_{{^3}He}\langle }1/2^+ m' | \sum_n \tau^+_n \bm{\sigma}_n 
|(1/2)^+ m{\rangle _{{^3}H}}.
\end{eqnarray}
Here, $m, m' = \pm 1/2$. 

Now we  sum up over the lepton polarizations, over the  projection of angular momenta, 
$m'$, of final nucleus and average over projection of angular momenta,  $m$,  of initial 
nucleus.  Evaluating the corresponding traces for 
the squared T-matrix we have
\begin{eqnarray}\label{T-mat}
\sum_{}~|\langle {f}|T^{(1)}|{i}\rangle |^2&=& 2~{\left(\frac{G_{\beta}}{~\sqrt{2}}\right)}^{{2}}
~F_0(Z+1,E_e) \left( B_F({^{3}H}) + B_{GT} ({^{3}H}) \right)
\end{eqnarray}
with
\begin{eqnarray}\label{MESQ}
B_F({^{3}H}) &=& |M_{F}|^{2} = \frac{1}{2} 
\sum_{m,m'} \left| M_F(m,m') \right|^2, \nonumber\\
B_{GT} ({^{3}H}) &=& g^2_A~|M_{GT}|^{2} = g^2_A~\frac{1}{2}
\sum_{m, m'} {\bm{M}}_{GT}(m,m')\cdot {\bm{M}}^*_{GT}(m,m').
\end{eqnarray}
We neglected terms proportional to the space components of the lepton momenta 
as they vanish after integration over angles in calculation of the total $\beta$-decay rate.

The Fermi and Gamow-Teller beta strengths in Eq. (\ref{MESQ}) can be represented in terms of 
reduced nuclear matrix elements:
\begin{eqnarray}\label{MF21}
B_F({^{3}H}) 
&=& \frac{1}{2}
\left| {_{{^3}He}\langle }(1/2)^+ \parallel  \sum_n \tau^+_n 
\parallel (1/2)^+ {\rangle _{{^3}H}} \right|^2, \\ 
\label{MGT21}
B_{GT}({^{3}H}) 
&=& g^2_A ~\frac{1}{2}
\left| {_{{^3}He}\langle }(1/2)^+ \parallel  \sum_n \tau^+_n \sigma_n
\parallel (1/2)^+{\rangle _{{^3}H}} \right|^2.
\end{eqnarray}

With our normalization of the lepton wave functions (\ref{WFS1e}), 
(\ref{WFnu}) the differential decay rate is
\begin{eqnarray}\label{DDR1}
d\Gamma^\beta = \sum |\langle f|T^{(1)}|i\rangle |^2 ~2\pi \delta(E_\nu + E_e + E_f - E_i)
~\frac{d {\bm{p}}_e}{(2\pi)^{3}}~ \frac{d{\bm{p}}_\nu}{(2\pi)^{3}}.
\end{eqnarray}  
Inserting Eq. (\ref{T-mat}) for squared T-matrix element 
and integrating over the phase space 
we find the total $\beta$-decay rate for the reaction (\ref{Tr-b})
\begin{eqnarray}\label{Total}
\Gamma^\beta({^{3}H}) = \frac{1}{2\pi^3} m_e \left( G_\beta m_e^2 \right)^2
\left( B_F({^{3}H}) + B_{GT} ({^{3}H}) \right)
~I^\beta ({^{3}H})
\end{eqnarray}  
with the phase space integral 
\begin{eqnarray}
\label{phase}
I^\beta({^{3}H}) = \frac{1}{m_e^5}
\int_{m_e}^{E_i-E_f} F_0(Z+1,E_e) p_e E_e (E_i-E_f-E_e)^2 dE_e.
\end{eqnarray}  
The dependence of $I^\beta ({^{3}H})$ on small neutrino mass is
ignored. For tritium numerical integration in (\ref{phase}) gives
\begin{equation}\label{num-I}
I^\beta({^{3}H}) = 2.88\times 10^{-6}.
\end{equation}
The nuclear matrix elements for tritium are known in the literature. We take for these matrix elements the values  
\mbox{$|M_{F}|^{2}=1, |M_{GT}|^{2}=3$} derived  in Ref. \cite{rasto}. Then we find
\begin{eqnarray}
T_{1/2}^\beta({^{3}H}) &=& \frac{\ln{2}}{\Gamma^\beta({^{3}H})} = 12.32~y,  
\end{eqnarray}  
which is very close to the measured value of tritium $\beta$-decay half-life. Vice versa, 
using Eqs. (\ref{Total})-(\ref{num-I}), one can extract a value of 
$\left|M_F\right|^2  + g^2_A \left|M_{GT}\right|^2$ in a model independent way from the experimental value 
$T_{1/2}^\beta(^{3}H)\approx 12.33\ y$. In this way we obtain 
\begin{eqnarray}\label{num-2}
\left|M_F\right|^2  + g^2_A \left|M_{GT}\right|^2 \approx 5.645.
\end{eqnarray}
This model independent estimate we will use in discussion  of  relic neutrino capture by tritium.

\subsection{$\beta$-decay of rhenium}
\label{sec12}

Single $\beta$-decay of rhenium 
\begin{equation}
{^{187}Re}((5/2)^+) \rightarrow {^{187}Os}((1/2)^-) + e^- + {\overline{\nu}}_e.  
\end{equation}
is more complicated than the tritium $\beta$-decay mainly because of  
the fact that the initial and final nuclear spins and parities are different. 
In \cite{dvor} it was shown that the dominant contribution to this decay 
is given by the emission
of electron and antineutrino in $p_{3/2}$ and $s_{1/2}$ wave states, respectively. 
The $p_{3/2}$ wave function of electron has the form 
\begin{eqnarray}\label{pw-Rh}
\Psi^{p_{3/2}} (\bm{x},E_e) 
&\simeq&  i ~\sqrt{F_1(Z+1,E_e)}~
\left( \bm{x}\cdot{\bm{p}}_e + \frac{1}{3}
 \bm\gamma\cdot\bm{x}
~\bm\gamma\cdot{\bm{p}}_e\right) ~ u(P_e).
\end{eqnarray}
$F_1$ is the relativistic Fermi  function (for more details see Appendix) \cite{doi}. 

Now, we are to analyze the squared matrix element in Eq. (\ref{ME1}) with the electron 
and neutrino wave functions given in 
Eqs.  (\ref{WFnu}), (\ref{pw-Rh}). Towards this end we consider the quantity  
\begin{eqnarray}
{\cal M} &=& 
{\bar{\psi}}_e (\bm{x},E_e) 
\gamma_{\mu} (1-\gamma_{{5}}) \psi^c_\nu (\bm{x},E_\nu )\times\nonumber\\ 
&& \langle (1/2)^- m' | \sum_n 
\tau^+_n ( g_V g^{\mu 0} +
g_A g^{\mu k} ({\bm\sigma}_n)^{k} )
| (5/2)^+ m \rangle~ \delta(\bm{x}-{\bm{x}}_n),
\end{eqnarray}
which is the integrand of the space integral in Eq. (\ref{ME1}). 
Carrying out summation over polarizations of leptons we keep only the terms
surviving after integration over phase space. Then, for 
the squared T-matrix element  we obtain \cite{Rhen}
\begin{eqnarray}
\sum~|\langle {f}|T^{(1)}|{i}\rangle |^2 &=& ~G_{\beta}^2 ~ g^2_A~\left|~ 
M^{(2^-)}_{GT} \right|^2 \frac{(p_e~R)^2}{3} F_1(Z+1,E_e).
\label{TSME2}
\end{eqnarray}
where the squared nuclear matrix element includes summation 
over all the spin orientations of the final nucleus $m'$ and 
averaging over the spin orientations of the initial nucleus $m$. 
The $\beta$-strength for ${^{187}Re}$ takes the form 
\begin{eqnarray}
B_{GT}({^{187}Re}) = g^2_A \left| M^{(2^-)}_{GT} \right|^2 
=  \frac{1}{6~R^2}~
\left| \langle (1/2)^- \parallel  \sum_n 
\tau^+_n 
\{ \bm{\sigma}_{n}^{~} \otimes \bm{r}_{n}^{~} \}_{2}
\parallel (5/2)^+\rangle \right|^2.
\end{eqnarray}

Inserting the above expression to Eq. (\ref{DDR1}) 
and integrating over the phase space of final electron and neutrino we find 
the total $\beta$-decay rate of rhenium
\begin{eqnarray}
\Gamma^\beta (^{187}Re) = ~m_e\frac{(G_{\beta} m_{e}^{2})^{2}}{2\pi^3} ~ \frac{(R m_{e})^{2}}{3}~
B_{GT}({^{187}Re}) ~I^\beta (^{187}Re),
\end{eqnarray}  
where the phase space integral is 
\begin{eqnarray}
\label{Rhe-I}
I^\beta (^{187}Re) &=& \frac{1}{m_e^7} \int_{m_e}^{E_i-E_f} 
p_e ~E_e ~(E_i - E_f -E_e)^2 F_{1}(Z+1,E_e) p^2_e dE_e. 
\end{eqnarray}  
We neglected the effect of small neutrino mass. 
For rhenium numerical integration in (\ref{Rhe-I}) gives
\begin{equation}
I^\beta(^{187}Re) = 1.22\times 10^{-7}.
\end{equation}
With this value of the phase space integral and 
using the experimental half-life
\begin{eqnarray}
T^\beta_{1/2} {(^{187}Re)}~=~4.35\times 10^{10}~years
\end{eqnarray}
we obtain the value of rhenium nuclear matrix element
\begin{eqnarray}\label{RhME}
g_A^2~\left|~M^{(2^-)}_{GT} \right|^2_{^{187}Re}= 3.57\times 10^{-4}.
\end{eqnarray}
We will use this value in our analysis of neutrino capture on 
rhenium where the same quantity will appear.

\section{Neutrino capture  with beta decaying  nuclei}
\label{sec2}

Now let us consider processes, similar to the previously considered $\beta$-decays, 
but with neutrino in the initial state of reaction:
\begin{equation}\label{nucapt1}
\nu_{e} + (A,Z) \rightarrow (A,Z+1)+ e^-.  
\end{equation}
This neutrino can be of any origin. An interesting possibility 
is a neutrino from the cosmic neutrino background.  
Thus the above reaction may represent one of the very few possible 
ways to detect this important component of the Universe. 
Since the energy of cosmic neutrinos is negligibly small, 
the above neutrino capture reaction takes place only if
there is no threshold.  
This implies that the target nucleus should be $\beta$-radioactive.  
In the previous sections  we considered in details two candidates of this type, tritium and 
rhenium. Below we study capture of cosmic neutrinos with these 
radioactive nuclei. 

The differential rate of the reaction (\ref{nucapt1})  is given by
\begin{eqnarray}\label{capt1}
d\Gamma^{\nu} = \sum \frac{1}{V} |\langle f|T^{(1)}|i\rangle |^2 
~2\pi \delta(E_e+E_f-E_i-E_\nu)~\frac{d {\bm{p}}_e}{(2\pi)^3}
\end{eqnarray}
for the normalization of incoming neutrino plane wave to be one particle per volume $V$.

The kinetic energy of electron produced in $\beta$-decay is 
continuously distributed within the interval 
$0 \leq  E_{e}-m_e \leq  Q_{\beta}-m_{\nu}$
while in the reaction (\ref{nucapt1}) the final electron is 
monoenergetic with $E_{e} - m_e = Q_{\beta} + E^{rel}_{\nu}$, where $E^{rel}_{\nu}$ 
is the relic neutrino energy.  Thus the signatures of $\beta$-decay and 
cosmic neutrino capture are quite different. However the energy 
difference between $\beta$-decay and neutrino capture emitted electrons 
is very small $\sim 2m_{\nu}$ which is challenging for energy resolution 
of detectors used in experiments searching for this process.  
>From the viewpoint of matrix element calculations neutrino 
capture reaction (\ref{nucapt1}) with very low energy neutrinos, such as 
the cosmic neutrinos, is very similar to $\beta$-decay of the target nucleus.

Below we consider two target nuclei tritium and rhenium which are planned to be used
in the experimental setups KATRIN with $^{3}$H and MARE with $^{187}$Re 
aimed to measure neutrino mass in sub-eV range. 

\subsection{Neutrino capture on tritium}
\label{sec21}

The cosmic neutrino capture reaction by tritium, 
\begin{equation}\label{Tr-cap}
{\nu}_e + ^{3}H((1/2)^+) \rightarrow ^{3}He((1/2)^+) + e^-, 
\end{equation}
is characterized by the T-matrix element of the same form as in the case
of single $\beta$-decay.  Within the same approximations and summing over the polarizations we have
\begin{eqnarray}\label{ME2CAP}
\sum~|\langle {f}|T^{(1)}|{i}\rangle |^2 
&=& G^2_{\beta}~ F_0(Z+1,E_e) 
\left( B_F({^{3}H}) + B_{GT} ({^{3}H}) \right).
\end{eqnarray}
Inserting  (\ref{ME2CAP}) to (\ref{capt1}), for the total capture rate we find
\begin{eqnarray}\label{TotG}
\Gamma^\nu ({^{3}H})  =  \frac{1}{V} \frac{1}{\pi} G^2_{\beta}
F_0(Z+1,E_e)
\left( B_F({^{3}H}) + B_{GT} ({^{3}H}) \right)~p_e~E_e,
\end{eqnarray}  
where $E_e = E_i + E_\nu - E_f$. 
With our neutrino wave function normalization both expressions correspond 
to capture rate per neutrino per $^{3}H$ atom.
For the number density of cosmic neutrinos $\eta_\nu$  we replace 
$1/V\rightarrow \eta_\nu$ in (\ref{TotG}) and get 
the capture rate per atom
\begin{eqnarray}\label{capt}
\Gamma^\nu ({^{3}H})  = 
\frac{1}{\pi} G_{\beta}^2 ~F_0(Z+1,E_e) ~p_e~E_e~
\left( B_F({^{3}H}) + B_{GT} ({^{3}H}) \right) 
~\frac{\eta_\nu}{\langle\eta_{\nu}\rangle}~\langle\eta_{\nu}\rangle.
\end{eqnarray}  
Here $\eta_\nu$ is the local cosmic neutrino number density which could be significantly 
larger than the average over the universe $\langle \eta_{\nu}\rangle \sim 56$ cm$^{-3}$ 
due to gravitational clustering \cite{vogel,clust}
\begin{eqnarray}\label{relden}
\frac{\eta_{\nu}}{\langle\eta_{\nu}\rangle} 
\sim 10^{3}-10^{4}
\end{eqnarray}
assuming $m_{\nu}=1$eV and baryon density $\eta_{b}=10^{-3}-10^{-4}$cm$^{-3}$ in a cluster of galaxies.

A combination of the nuclear matrix elements in Eq. (\ref{capt}), 
shown in curl  brackets, we previously encountered in $\beta$-decay 
rate and found its numerical value from the half-life $T_{1/2}$ of 
$^{3}$H. The corresponding value is given in Eq. (\ref{num-2}) which 
we substitute in (\ref{capt}) and obtain
 \begin{eqnarray}\label{rateval}
 \Gamma^\nu (^{3}H)= 4.2\times 10^{-25} \frac{\eta_\nu}{\langle\eta_{\nu}\rangle} ~y^{-1}.
 \end{eqnarray}
This value is compatible with the result of Ref. \cite{vogel, cocco} for the case 
$\eta_\nu= \langle\eta_{\nu}\rangle$.

The KATRIN experiment, dedicated to measurement of electron neutrino mass  from 
endpoint  electron energy spectrum of $\beta$-decay of tritium, can, in principle, also search for cosmic neutrinos
via neutrino capture reaction (\ref{Tr-cap}).  The KATRIN experiment (in construction phase) aims to measure  $m_\nu$ with sensitivity of 0.2 eV \cite{otten} using about $50~\mu g$ of tritium
corresponding to $5\times 10^{18}$ $T_2$ molecules \cite{clust-det}.  For this amount of target nuclei we find from Eq. (\ref{rateval}) 
the number of neutrino capture events
\begin{equation}
\label{KATRIN-tot}
N^{\nu}_{capt}(KATRIN) \approx 4.2 \times 10^{-6}~\frac{\eta_\nu}{\langle\eta_{\nu}\rangle}
~y^{-1}.
\end{equation}
Considering this estimate we conclude that observation of
relic neutrino capture in KATRIN experiment looks rather unrealistic.  Further comments on this subject will be given in 
sec. \ref{conclusions}.

\subsection{Neutrino capture on rhenium}
\label{sec22}

The cosmic neutrino capture by rhenium
\begin{equation}
{\nu}_e + ^{187}Re((5/2)^+) \rightarrow {^{187}Os}((1/2)^-) + e^- 
\end{equation}
we analyze in the way similar to the previous case of tritium. 
The T-matrix determining this process coincide with the rhenium $\beta$-decay 
matrix element shown in Eq. (\ref{TSME2}).
After neglecting the part of the electron wave function
associated with emission of the $s_{1/2}$-electron, which
is very small, the squared and  summed over polarizations  T-matrix 
takes the form
\begin{eqnarray}
\sum~|\langle {f}|T^{(1)}|{i}\rangle |^2 
&=&  G_{\beta}^2 ~ B_{GT}({^{187}Re})~\frac{(p_e~R)^2}{3}~
 F_1(Z+1,E_e).
\end{eqnarray}
Inserting this expression in Eq. (\ref{capt1}), for 
the total  capture rate per cosmic neutrino we obtain
\begin{eqnarray}
\Gamma^\nu (^{187}Re) = \frac{1}{V}~ \frac{1}{\pi}~ G^2_\beta
~ B_{GT}({^{187}Re})~
F_1(Z+1,E_e)~\frac{(p_e~R)^2}{3}
~p_e~E_e
\end{eqnarray}  
with $E_e = E_i + E_\nu - E_f$. 

As in the case of tritium we replace $1/V$ with the neutrino number density $\eta_\nu$ and get
the capture rate per atom of rhenium
\begin{eqnarray}
\Gamma^\nu (^{187}Re)  =
\frac{1}{\pi} G^2_\beta ~F_1(Z+1,E_e)~
\frac{(p_e~ R)^2}{3}~B_{GT}({^{187}Re})~
p_e~E_e~
~\frac{\eta_\nu}{\langle\eta_{\nu}\rangle}~\langle\eta_{\nu}\rangle.
\end{eqnarray}
Substituting to this equation numerical values of nuclear matrix element 
from (\ref{RhME}) and other constants we obtain  
for the capture rate per atom of $^{187}Re$ 
\begin{eqnarray}
\Gamma^\nu (^{187}Re) =  2.75\times 10^{-32}~\frac{\eta_\nu}{\langle\eta_{\nu}\rangle}~y^{-1}. 
\end{eqnarray}

For a detector with 100 g of $^{187}Re$, i.e. with  $3.2\times 10^{23}$ rhenium atoms, 
we find the number of cosmic neutrino capture events
\begin{equation}
N^\nu_{capt} \simeq 
8.9\times 10^{-9} \frac{\eta_\nu}{\langle\eta_{\nu}\rangle}~y^{-1}. 
\end{equation}

The MARE project will investigate the $\beta$-decay of $^{187}Re$ 
with absorbers of metallic rhenium or $AgReO_4$. It foresees a 760 grams 
bolometer. For this amount of rhenium  the number of neutrino capture
events is 
\begin{eqnarray}
\label{MARE-tot}
N^{\nu}_{capt} (MARE) \simeq
6.7\times 10^{-8}~\frac{\eta_\nu}{\langle\eta_{\nu}\rangle}
~y^{-1}.
\end{eqnarray}
The MARE detector technology can, in principle, be scaled up.  With about 4 orders more rhenium 
the capture rate would become about two orders of magnitude larger than in case of 
the KATRIN experiment.
Note that the  KATRIN experiment can hardly be significantly scaled up due 
to safety limitations on the amount of  highly radioactive tritium.

\section{Discussions and Conclusions}
\label{conclusions}

As was shown in the previous sections both with $^{3}$H and $^{187}$Re target nuclei the relic neutrino capture rate Eqs. (\ref{KATRIN-tot})-(\ref{MARE-tot}) event numbers are extremely small and unobservable in the present and near future experiments. 
The gravitational clustering of relic neutrinos at the level of ${\eta_\nu}/{\langle\eta_{\nu}\rangle} \simeq 10^{3}-10^{4}$ \cite{vogel,clust}  can hardly change this conclusion. On the other hand this effect 
could be significantly stronger leading to a clustering about 
${\eta_\nu}/{\langle\eta_{\nu}\rangle} \simeq 10^{13}$ discussed in Ref. \cite{superclust}. Then the capture rate 
becomes large reaching to  about $N^{\nu}_{capt} \sim 10^{7}~y^{-1}$ which looks very promising.
However, a key point is distinguishing the relic neutrino capture signal from the $\beta$-decay background.
In this respect the main quantity to be analyzed is the ratio  $\lambda_{\nu}/\lambda_{\beta}$ of the the partial rates of $\beta$-decay$\lambda_{\beta}$ and  relic neutrino capture $\lambda_{\nu}$ with the electrons in the energy interval around the endpoint  $Q_{\beta}$ of the $\beta$-decay electron energy spectrum. Despite the 
$\beta$-decay spectrum ends at $Q_{\beta}-m_{\nu}$ and the electrons from the neutrino capture locates at $Q_{\beta}+m_{\nu}$ in a realistic experiment they overlap due to finite energy resolution $\Delta$ of a detector. As it was shown in Refs. \cite{cocco,vogel} the neutrino capture events can be discriminated from the 
$\beta$-decay background if $m_{\nu}/\Delta \sim 2$ or smaller. This condition is very challenging for experiments and is nearly independent of target nucleus and very weakly depend on the clustering 
$\eta_{\nu}/\langle \eta_{\nu}\rangle$. 
Only a strong clustering $\eta_{\nu}/\langle \eta_{\nu}\rangle \sim 10^{13}$, considered in Ref. \cite{superclust}, may relax this condition and simultaneously significantly increase the total event rate of the relic neutrino capture. We note also that the considered searches for relic neutrinos via neutrino capture on $\beta$-decaying nuclei even been unable to detect this process may set useful limits on the actual relic neutrino density 
$\eta_{\nu}/\langle \eta_{\nu}\rangle$ in the Earth vicinity.  These possibility requires additional study.  However it should be pointed out that there is a big uncertainty in this sort of studies due to unknown absolute mass scale of neutrino. If the forthcoming 
$\beta$-decay experiments  \cite{katrin,otten},  \cite{mare} set the value of $m_{\nu_{e}}$ within their sensitivities then the energy resolution necessary for successful searches of relic neutrinos will be within reasonable values like $\Delta \sim 0.1$eV \cite{cocco}, which could be achieved in future experiments.

In conclusion,
we carried out a detailed analysis of nuclear physics and kinematical aspects of 
single nuclear $\beta$-decay and relic neutrino capture on $\beta$-radioactive nuclei.
We focussed on $^{3}$H and $^{187}$Re isotopes to be used in the forthcoming  KATRIN and in the planned MARE experiments, respectively.

As to the prospects of relic neutrino direct detection through neutrino capture on  $\beta$-decaying nuclei
we concluded that they are rather pessimistic with the present experimental methods and techniques.
Even taking into account the gravitational clustering, which greatly enhances the relic neutrino number density in galaxy clusters,
does not improve this situation: the detection rate remains very small and hardly observable in the near future. 
However we noted that scaling MARE experiment up to several hundreds of kilograms of rhenium
would offer an event rate significantly larger than in KATRIN experiment. 
This possibility, however, remains  technically very questionable.  

Despite our rather pessimistic conclusions we point out that searches for cosmic neutrino 
capture via beta decaying nuclei deserves to be carried out. After all this is the only 
known direct way to probe this component of the universe and from non-observation of 
the capture process to extract limits on the local neutrino number density $\eta_{\nu}$,
which could be complementary to the known cosmological and astrophysical limits. 
On the other hand one can never exclude that we are passing through a neutrino clump 
with a density much higher than predicted by the gravitational clustering \cite{vogel,clust, superclust}. 
This would eventually make observation of cosmic neutrino capture realistic in 
even the forthcoming experiments.

\begin{acknowledgments}
Authors are grateful to V. Egorov, S. Bilenky and Yu. Kamyshkov for  stimulating discussions.
F.\v S. and R.H. acknowledge the support of the VEGA Grant 
agency of the Slovak Republic under the Contract No. 1/0639/09.
S.K.  acknowledges the support of  \mbox{FONDECYT} grant 1100582, and 
Centro Cient\'{\i}fico-Tecnol\'ogico de Valpara\'iso PBCT ACT-028.  
\end{acknowledgments}

\appendix*\section{Partial electron wave functions}

Here, for convenience, we show several terms of the partial 
wave decomposition of the relativistic electron wave function 
in the Coulomb field of a uniform charge distribution of a nucleus:
\begin{equation}
\Psi_e (\bm{r},E) = 
\Psi^{s_{1/2}}(\bm{r},E) + \Psi^{p_{1/2}}(\bm{r},E) +
\Psi^{p_{3/2}}(\bm{r},E) + ...
\end{equation}
In our notation we follow Ref.  \cite{doi}. Keeping the leading 
terms in r we have for the partial wave functions:

$s_{1/2}$  wave:
\begin{eqnarray}
\Psi^{s_{1/2}}_s(\bm{r},E) &=& 
\left( \begin{array}{c}
{\tilde{g}}_{-1}~ \chi_s  \\
{\tilde{f}}_{+1}~\bm\sigma\cdot\hat{\bm{p}}~ 
 \chi_s 
\end{array} \right) \nonumber\\
&\simeq& \sqrt{F_0(Z,E)}~ u_s(P).
\end{eqnarray}

$p_{1/2}$ wave: 
\begin{eqnarray}
\Psi^{p_{1/2}}_s(\bm{r},E) &=&  
i~  \left( \begin{array}{c}
{\tilde{g}}_{+1}~\bm\sigma\cdot\hat{\bm{r}}~
\bm\sigma\cdot\hat{\bm{p}}~ \chi_s \\
-{\tilde{f}}_{-1}~ 
\bm\sigma\cdot\hat{\bm{r}}~\chi_s 
\end{array} \right) \nonumber\\
&\simeq& 
i ~\frac{\alpha Z}{2}~\sqrt{F_0(Z,E)}~ 
\gamma_0~\bm{\gamma}\cdot\hat{\bm{r}} ~u_s(P).
\end{eqnarray}

$p_{3/2}$ wave:
\begin{eqnarray}
\Psi^{p_{3/2}}_s(\bm{r},E) &=&
i~\left( \begin{array}{c}
{\tilde{g}}_{-2}~\left[3 \hat{\bm{r}}\cdot\hat{\bm{p}} -
\bm\sigma\cdot\hat{\bm{r}}~\bm\sigma\cdot\hat{\bm{p}} \right]~ \chi_s \\
{\tilde{f}}_{+2}~\left[ 
3 \hat{\bm{r}}\cdot\hat{\bm{p}}~\bm\sigma\cdot\hat{\bm{p}} 
- \bm\sigma\cdot\hat{\bm{r}}\right]~\chi_s 
\end{array} \right) \nonumber \\
&\simeq&  i ~\sqrt{F_1(Z,E)}~
\left( \bm{r}\cdot\bm{p} + \frac{1}{3}
 \bm\gamma\cdot\bm{r}
~\bm\gamma\cdot\bm{p}\right) ~ u_s(P).
\end{eqnarray}
The relativistic Fermi function $F_0(Z,E_e)$ ($F_1(Z,E_e)$) 
takes into account the Coulomb interaction of emitted 
$s_{1/2}$ and $p_{1/2}$ ($p_{3/2}$) electrons 
with the nucleus and is given, for instance, in Ref.  \cite{doi}. 
The Dirac spinors are normalized as $u^{\dagger}(P)u(P)=1$.

\end{document}